\documentclass[showpacs,preprintnumbers,amsmath,amssymb]{revtex4}
\usepackage{amsmath,amssymb,graphics,epsfig,subfigure}
\usepackage{color}

\begin{document}
\newcommand {\nn} {\nonumber}
\renewcommand{\baselinestretch}{1.3}

\title{Clapeyron equations and fitting formula of the coexistence curve in the extended phase space of charged AdS black holes}

\author{Shao-Wen Wei \footnote{weishw@lzu.edu.cn},
        Yu-Xiao Liu \footnote{liuyx@lzu.edu.cn}}

\affiliation{Institute of Theoretical Physics, Lanzhou University, Lanzhou 730000, People's Republic of China}

\begin{abstract}
In this paper, we first review the equal area laws and Clapeyron equations in the extended phase space of the charged anti-de Sitter black holes. With different fixed parameters, the Maxwell's equal area law holds not only in the pressure-thermodynamic volume oscillatory line, but also in the charge-electric potential and temperature-entropy oscillatory lines. The conventional Clapeyron equation is generalized and two extra equations are found. Moreover, we show that the coexistence curve of the small and large charged black holes is charge independent in the reduced parameter space for any dimension of spacetime. The highly accurate fitting formula for the coexistence curve is also presented. Using this fitting formula of the coexistence curve, we find that the Clapeyron equations are highly consistent with the calculated values. The fitting formula is also very useful for further study on the thermodynamic property of the system varying along the coexistence curve.
\end{abstract}



\maketitle

\section{Introduction}
\label{secIntroduction}

Black hole thermodynamics has been an exciting and challenging topic in theoretical physics. Inspired especially by the AdS/CFT correspondence \cite{Maldacena,Gubser,Witten}, the anti-de Sitter (AdS) black hole thermodynamics has attracted much attention. It was first noted that in AdS space, there exists Hawking-Page phase transition between a stable large black hole and thermal gas \cite{Hawking}, which can be explained as the confinement/deconfinement phase transition of the gauge field \cite{Hawking,Witten2}.

Further study of charged AdS black holes revealed that there is a first-order phase transition between small and large black holes in the canonical ensemble \cite{Chamblin,Chamblin2,Roychowdhury,Banerjee}. With an increase of the temperature, the phase transition point terminates at a critical point, where the first-order phase transition becomes a second-order one. The behavior of such a phase transition is quite similar to the van der Waals (vdW) fluid.

Most recently, the investigation of the AdS black hole thermodynamics has been generalized to the extended phase space, where the cosmological constant is treated as a dynamical pressure and its conjugate quantity is the thermodynamic volume of the black hole \cite{Dolan,Cvetic,Kastor,Traschen,Castro,Menoufi}. Phase transition in the charged AdS black hole was reexamined in Ref. \cite{Kubiznak}. It was found that the black hole system and the vdW system share the same pressure-thermodynamic ($P-V$)oscillatory behavior, critical exponents, and scaling laws. Thus, the precise analogy between the charged AdS black hole and the vdW system was established. Later, the study of the AdS black hole phase transition in extended phase space was generalized to different AdS black hole backgrounds \cite{Gunasekaran,Hendi,Chen,ZhaoZhao,Altamirano,Cai,AltamiranoKubiznak,XuXu,Mo,zou,MoLiu,
Altamirano3,Wei,Wei2,Zhang,Moliu,Zou2,Zhao2,ZhaoZhang,XuZhang,Frassino,Zhangcai,Mirza,Kostouki,Rajagopal,hhzhao}. In these studies, it was shown that besides the small/large black hole first-order phase transition reminiscent of the liquid/gas transition of the vdW fluid, there were also some new interesting phenomena, such as reentrant phase transitions of multicomponent liquids, multiple first-order solid/liquid/gas phase transitions, and liquid/gas phase transitions of the vdW type. For example, the ($d\geq 6$)-dimensional single spinning vacuum Kerr-AdS black hole demonstrates the peculiar behavior of the large/small/large black hole phase transitions \cite{Altamirano}, and the multiply rotating Kerr-AdS black hole system displays a small/intermediate/large black hole phase transition with one tricritical and two critical points in some range of the parameters \cite{AltamiranoKubiznak}. The doubly spinning Myers-Perry black hole also has a similar reentrant phase transition structure \cite{Altamirano3}. Such a reentrant phase transition is also observed in modified gravity theories, such as Born-Infeld gravity \cite{Gunasekaran}, Gauss-Bonnet gravity \cite{Wei2}, and Lovelock gravity \cite{Frassino}.

On the other hand, understanding the thermodynamic properties of a black hole system along or across the coexistence curve is also an interesting subject. The authors of Ref. \cite{Liuwang} calculated the quasinormal modes of massless scalar perturbations around the small and large four-dimensional charged AdS black holes. The result showed that the quasinormal modes can be a dynamic probe of the thermodynamic phase transition.

Toward understanding the thermodynamic properties of the charged AdS black hole system along or across the coexistence curve, one expects an analytic formula of the coexistence curve. However, there is only a numerical calculation of it. In this paper, we reconsider the small/large black hole phase transition and present the fitting formula of the coexistence curve in the reduced parameter space for ($d=4\sim10$)-dimensional charged AdS black holes. The result confirms our analysis that the fitting formula is independent of charge for any dimension $d$. Along the coexistence curve, the Clapeyron equations are also reexamined. Besides the conventional one, two additional equations are obtained in the extended phase space. Based on the fitting formula of the coexistence curve, these Clapeyron equations are numerically checked.

This paper is organized as follows. In Sec. \ref{dddd}, we investigate the equal area laws and the Clapeyron equations in the extended phase space for the charged AdS black hole. In Sec. \ref{equalarealaw}, we present the fitting formula of the coexistence curve and check the Clapeyron equations. We will summarize our results in Sec. \ref{Conclusion}.

\section{Equal area laws and Clapeyron equations}
\label{dddd}

Recently, there has been great interest in studying the phase transition by treating the cosmological constant as a dynamical pressure, i.e.,
\begin{eqnarray}
 P=-\frac{\Lambda}{8\pi}=\frac{(d-1)(d-2)}{16\pi l^{2}}.
\end{eqnarray}
With such a new interpretation, the cosmological constant can be regarded as a new thermodynamic variable. In this extended phase space, the black hole mass $M$ is considered as the enthalpy $H\equiv M$ rather than the internal energy of the gravitational system \cite{KastorRay}, and the Smarr relation for a $d$-dimensional charged AdS black hole is modified as
\begin{eqnarray}
 (d-3)H=(d-2)TS-2PV+(d-3)Q\Phi.\label{mass}
\end{eqnarray}
Differentiating it, we get the first law:
\begin{eqnarray}
 dH=TdS+VdP+\Phi dQ,\label{firstlaw}
\end{eqnarray}
with $V$ being the thermodynamic volume
\begin{eqnarray}
 V=\bigg(\frac{\partial H}{\partial P}\bigg)_{S,Q}.
\end{eqnarray}
Furthermore, the Gibbs free energy can be obtained by the Legendre transformations
\begin{eqnarray}
 G&=&H-TS,\\
 dG&=&-SdT+\Phi dQ+VdP.\label{Gibss}
\end{eqnarray}
For a specific spacetime background, one can clearly get its Gibbs free energy. Considering that the thermodynamic system prefers a low Gibbs free energy, we then can obtain its phase transition information. Now the study of phase transition has been applied to a number of AdS black holes, and richer phase structures are displayed. Many AdS black hole systems exhibit the small/large black hole phase transition, which is reminiscent of the vdW liquid/gas phase transition. The state equation for them can be expressed as
\begin{eqnarray}
 P=\frac{T}{v}+\mathcal{O}\bigg(\frac{1}{v^{2}}\bigg).
\end{eqnarray}
Analogous to the vdW fluid, here $v$ can be interpreted as the specific volume of the black hole fluid. Study shows that both of them share the similar physical properties, such as the $P-v$ (pressure-specific volume) oscillatory behavior, the critical exponents, and scaling laws near the critical point. Moreover the reentrant phase transitions were also found in the black hole system.

In general, the phase diagram or the coexistence curve can be obtained by analyzing the characteristic swallow tail behavior of the Gibbs free energy, and it is extensively used in the related papers. On the other hand, it can also be obtained by constructing the Maxwell's equal area law. However, after a simple check, we arrive at the result that the phase diagrams obtained from the Gibbs free energy and the equal area law in $P-v$ behavior are not consistent with each other. We suggest that the problem originates from the confusion of thermodynamic volume $V$ and specific volume $v$. For the vdW fluid, the equal area law is effective for both thermodynamic and specific volumes, because $V=Nv$, with $N$ being a fixed constant denoting the total number of the fluid molecules. For the black hole case, however, such a linear relation does not hold anymore. Thus, considering Eq. (\ref{Gibss}), the Maxwell's equal area law is effective only for the thermodynamic volume $V$ in a black hole system.

\subsection{Equal area law}

\begin{figure}
\includegraphics[width=8cm,height=6cm]{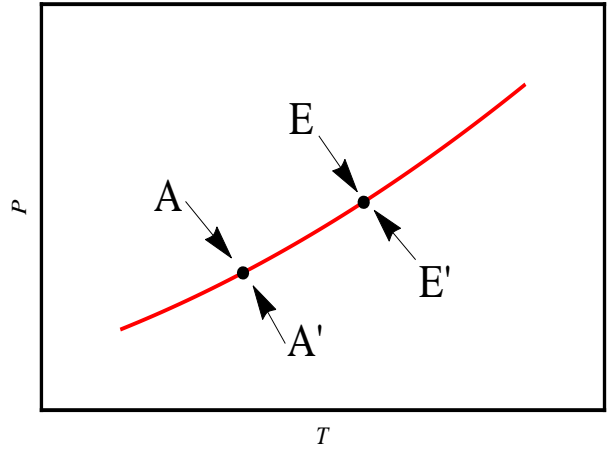}
\caption{Sketch picture of the coexistence curve. States $A$ and $A'$, $E$ and $E'$, respectively, can coexist with each other.}\label{pt0}
\end{figure}

Before discussing the equal area law, we plot a sketch picture of the phase diagram in Fig. \ref{pt0}. The red line denotes the coexistence curve, above and below which there are two different system phases. Each point located on the curve corresponds to two different phases while having the same Gibbs free energy. For example, states $A$ and $E$ are related by one system phase, while states $A'$ and $E'$ by another phase. Between them, states $A$ $(E)$ and $A'$ $(E')$ have the same Gibbs free energy.

Considering that a system transforms state $A$ to $A'$, i.e., $G_{A}=G_{A'}$, we have
\begin{eqnarray}
 -SdT+\Phi dQ+VdP =0.
\end{eqnarray}
Now, we integrate it:
\begin{eqnarray}
 -\int^{T_{A'}}_{T_{A}} SdT+\int^{Q_{A'}}_{Q_{A}} \Phi dQ+\int^{P_{A'}}_{P_{A}} VdP =0.\label{int}
\end{eqnarray}

\begin{figure}
\center{\subfigure[]{\label{PVT}
\includegraphics[width=6cm,height=6cm]{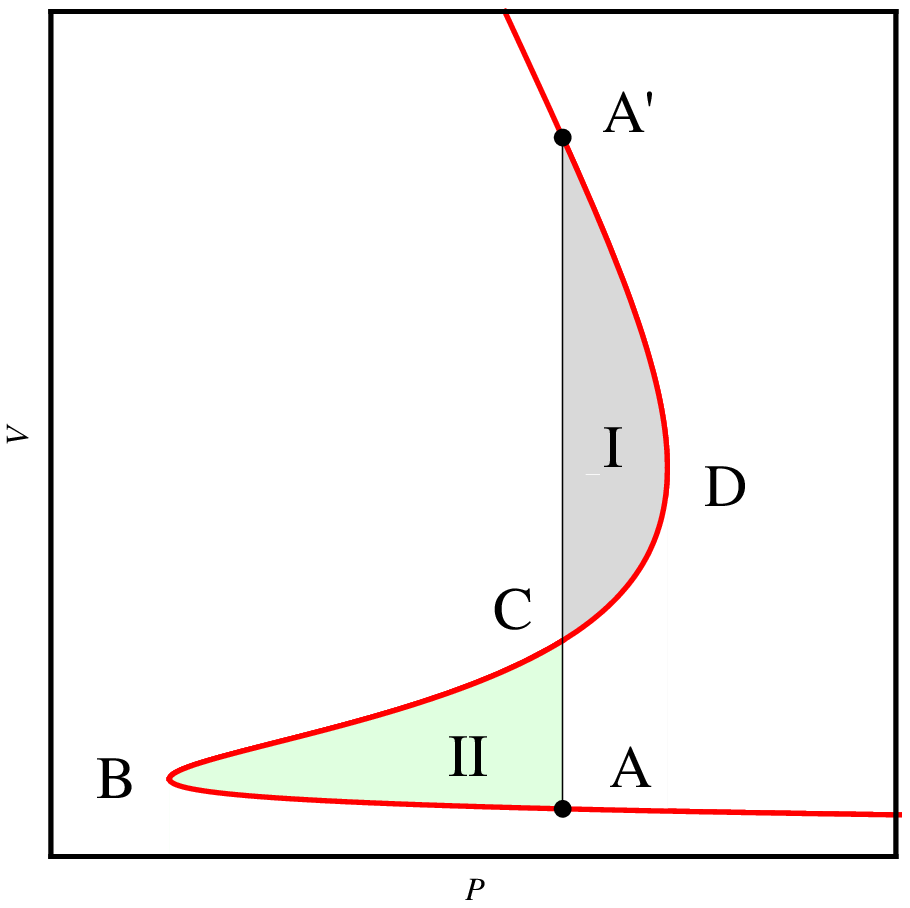}}
\subfigure[]{\label{PTV}
\includegraphics[width=6cm,height=6cm]{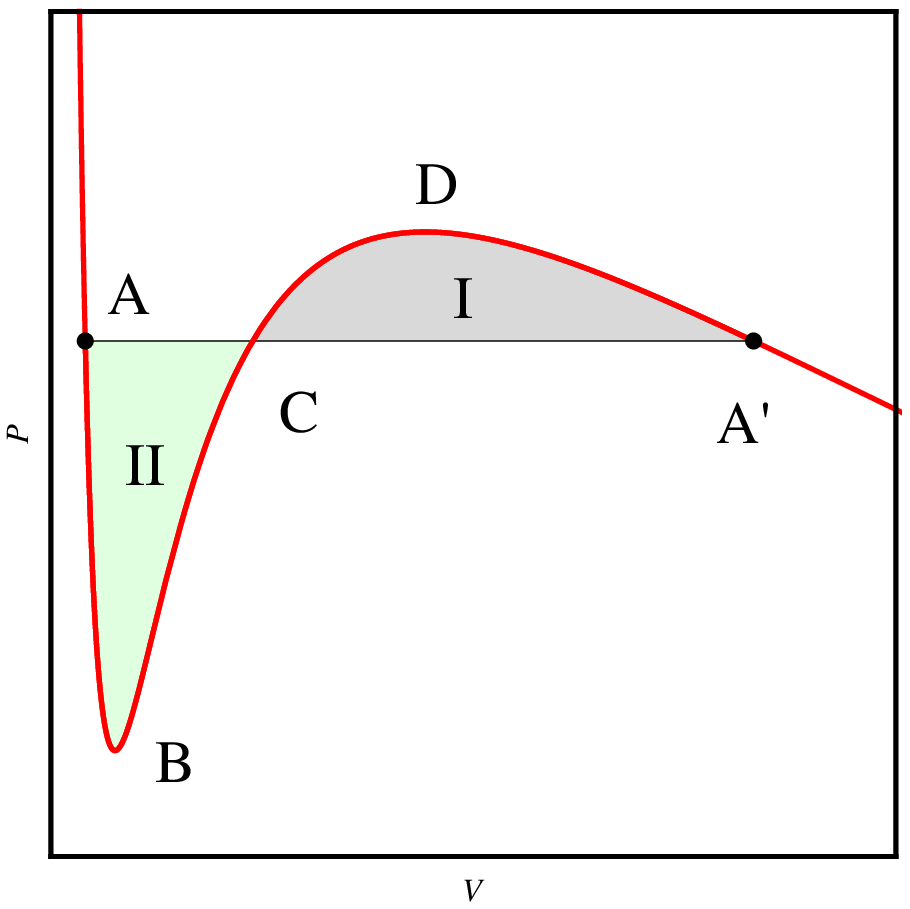}}}
\caption{$P-V$ oscillatory behavior. (a) $V-P$ plane. (b) $P-V$ plane.}
\end{figure}

\noindent\textbf{Case I:} charge $Q$ and temperature $T$ fixed.
For this case, we consider the isotherm in the $P-V$ plane. Therefore, the first and second terms in Eq. (\ref{int}) vanish, and we get
\begin{eqnarray}
 \int^{P_{A'}}_{P_{A}} VdP =0.\label{vp0}
\end{eqnarray}
Along the coexistence curve, we know that the two states have the same pressure, i.e., $P_{A'}=P_{A}$. Here we show the isotherm in the $V-P$ plane in Fig. \ref{PVT}. It is clear that oscillatory behavior exists. Then the integral in Eq. (\ref{vp0}) can be reexpressed as
\begin{eqnarray}
 \bigg(\int^{P_{B}}_{P_{A}} VdP+\int^{P_{C}}_{P_{B}} VdP\bigg)+
  \bigg(\int^{P_{D}}_{P_{C}} VdP+\int^{P_{A'}}_{P_{D}} VdP\bigg)  =0. \label{interalVdP}
\end{eqnarray}
The pressure $P_{C}=P_{A}=P_{A'}$ is equivalent to that of the phase transition pressure. From Fig. \ref{PVT}, we are see clearly that the first bracket in (\ref{interalVdP}) denotes the area of region II and the second one the negative area of region I. Thus, we get
\begin{eqnarray}
 \text{Area(I)}=\text{Area(II)}.
\end{eqnarray}
This is in fact the Maxwell's equal area law, and we are familiar with it as showed in the $P-V$ plane [see Fig. \ref{PTV}]. Therefore, we see that this equal area law holds during the phase transition. On the other hand, we can also use such an area law to find the value of the phase transition parameter. Letting the temperature $T$ take all possible values, we will obtain the phase diagram for each charge $Q$ using such an area law. It is also worthwhile to note that this equal area law is effective for the thermodynamic volume $V$ rather the specific volume $v$.

%%%%%%%%%%%%%%%%%%%%%%%%%%%%%%%%%%%%%%%%%%%%%%%%%%%%%%%%%%%%%%%%%%%%%
%%%%%%%%%%%%%%%%%%%%%%%%%%%%%%%%%%%%%%%%%%%%%%%%%%%%%%%%%%%%%%%%%%%%%%%%%

%%%%%%%%%%%%%%%%%%%%%%%%%%%%%%%%%%%%%%%%%%%%%%%%%%%%%%%%%%%%%%%%%%%%%
%%%%%%%%%%%%%%%%%%%%%%%%%%%%%%%%%%%%%%%%%%%%%%%%%%%%%%%%%%%%%%%%%%%%%%%%%

\noindent\textbf{Case II:} charge $Q$ and pressure $P$ fixed.
In the above case, we see that the equal area law holds in the $P-V$ plane with $Q$ and $T$ fixed. There is an equivalent choice in which $Q$ and $P$ are fixed. Under such a choice, one can carry out the above analysis and the equal area law will be found in the temperature-entropy ($T-S$) plane. This generalized area law was first studied in Refs. \cite{Spallucci,Smailagic}.

\noindent\textbf{Case III:} pressure $P$ and temperature $T$ fixed. During the phase transition, Eq. (\ref{int}) reduces to
\begin{eqnarray}
 \int^{Q_{A'}}_{Q_{A}} \Phi dQ=0,
\end{eqnarray}
which can be reexpressed as
\begin{eqnarray}
 \int_{\Phi_{A'}}^{\Phi_{A}}Q(\Phi)d\Phi=(\Phi_{A'}-\Phi_{A})Q(\Phi_{A}),
\end{eqnarray}
where $Q_{A}=Q_{A'}$ is considered. This equation is just another expression of the equal area law.

Here, it is clear that the Maxwell's equal area law for the charged AdS black hole gets a generalization. And the three area laws are consistent with each other.

\subsection{Phase transition pressure}

Here we continue to consider the manner in which the black hole system passes from state $A$ to $A'$. The Gibbs free energies related to the two states are
\begin{eqnarray}
 G_{A}&=&\frac{1}{(d-3)}T_{A}S_{A}+\Phi_{A}Q_{A}-\frac{2}{(d-3)}P_{A}V_{A},\\
 G_{A'}&=&\frac{1}{(d-3)}T_{A'}S_{A'}+\Phi_{A'}Q_{A'}-\frac{2}{(d-3)}P_{A'}V_{A'}.
\end{eqnarray}
Taking charge $Q$ as fixed and noting that $T_{A}=T_{A'}, P_{A}=P_{A'}$, we have
\begin{eqnarray}
 \frac{(S_{A}-S_{A'})}{(d-3)}T_{A}+(\Phi_{A}-\Phi_{A'})Q-\frac{2}{(d-3)}(V_{A}-V_{A'})P_{A}=0.
\end{eqnarray}
Then the phase transition pressure can be expressed as
\begin{eqnarray}
 P_{A}=\frac{(d-3)Q}{2}\frac{\Delta \Phi}{\Delta V}
     +\frac{T_{A}}{2}\frac{\Delta S}{\Delta V}.\label{PA}
\end{eqnarray}
A similar calculation can be applied to the case of fixed $\Phi$, which corresponds to the grand canonical ensemble, and Eq. (\ref{PA}) becomes
\begin{eqnarray}
 P_{A}=\frac{(d-3)\Phi}{2}\frac{\Delta Q}{\Delta V}
     +\frac{T_{A}}{2}\frac{\Delta S}{\Delta V}.
\end{eqnarray}

\subsection{Clapeyron equations}

The Clapeyron equation is a useful equation for studying the liquid/gas phase transition of the vdW fluid. Here we will review it and make a generalization to the charged black hole system among the small/large black hole phase transition.

Since states $A$ and $A'$, $E$ and $E'$ are in phase equilibrium,
\begin{eqnarray}
 G_{A}=G_{A'}, G_{E}=G_{E'}.
\end{eqnarray}
Thus
\begin{eqnarray}
 G_{E}-G_{A}=G_{E'}-G_{A'}.\label{GG}
\end{eqnarray}
If states $A$ ($A'$) and $E$ ($E'$) are very close, we then have
\begin{eqnarray}
 G_{E}-G_{A}=dG&=&-S_{A}dT+\Phi_{A} dQ+V_{A}dP,\\
 G_{E'}-G_{A'}=dG&=&-S_{A'}dT+\Phi_{A'} dQ+V_{A'}dP.
\end{eqnarray}
By inserting them into Eq. (\ref{GG}), we easily get
\begin{eqnarray}
 -S_{A'}dT+\Phi_{A'} dQ+V_{A'}dP=-S_{A}dT+\Phi_{A} dQ+V_{A}dP.
\end{eqnarray}
Taking $Q$, $P$, and $T$ as fixed, we obtain, respectively
\begin{eqnarray}
 \bigg(\frac{dP}{dT}\bigg)_{Q}&=&\frac{S_{A'}-S_{A}}{V_{A'}-V_{A}}=\frac{\Delta S}{\Delta V},\label{Clapeyron}\\
 \bigg(\frac{dQ}{dP}\bigg)_{T}&=&\frac{V_{A'}-V_{A}}{\Phi_{A'}-\Phi_{A}}=-\frac{\Delta V}{\Delta \Phi},\label{Clapeyron2}\\
 \bigg(\frac{dT}{dQ}\bigg)_{P}&=&\frac{\Phi_{A'}-\Phi_{A}}{S_{A'}-S_{A}}=\frac{\Delta \Phi}{\Delta S}.\label{Clapeyron3}
\end{eqnarray}
The first one is just the conventional Clapeyron equation. The other two are the generalized Clapeyron equations in the extended phase space of the charged AdS black hole system. In Ref. \cite{Dolan3}, the author has checked Eq. (\ref{Clapeyron}) for the black hole/AdS space phase transition in the nonrotating and rotating black hole systems. However, checking these equations during the small/large black hole phase transition requires a numerical calculation, and we will carry that out in the next section. Using these equations, one can easily check the Maxwell's relation $\big(\frac{dP}{dT}\big)_{Q}\big(\frac{dT}{dQ}\big)_{P}\big(\frac{dQ}{dP}\big)_{T}=-1$. On the other hand, the pressure $P_{A}$ in Eq. (\ref{PA}) can be reexpressed as
\begin{eqnarray}
 P_{A}=-\frac{(d-3)Q}{2}\bigg(\frac{dP}{dQ}\bigg)_{T}
      +\frac{T_{A}}{2}\bigg(\frac{dP}{dT}\bigg)_{Q}.\label{Clapeyron4}
\end{eqnarray}

\section{Fitting formula of the coexistence curve}
\label{equalarealaw}

\begin{figure}
\includegraphics[width=8cm,height=6cm]{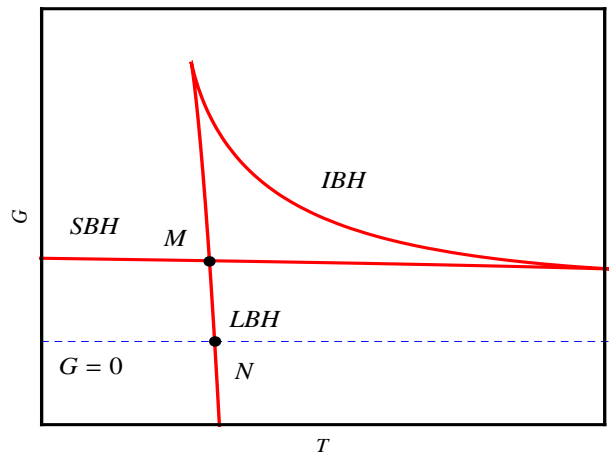}
\caption{Phase transition and characteristic swallow tail behavior of the Gibbs free energy. SBH, IBH, and LBH denote small, intermediate, and large black holes, respectively.}\label{Pgtq}
\end{figure}

Here we first give a brief review of the charged AdS black hole. Its line element is given by
\begin{eqnarray}
 ds^{2}&=&-f(r)dt^{2}+f^{-1}(r)dr^{2}+r^{2}d\Omega^{2}_{d-2},\\
 f(r)&=&1-\frac{m}{r^{d-3}}+\frac{q^{2}}{r^{2(d-3)}}+\frac{r^{2}}{l^{2}}.
\end{eqnarray}
The parameters $m$ and $q$, relate to the enthalpy and charge of the black hole, respectively
\begin{eqnarray}
 H&=&\frac{d-2}{16\pi}\omega_{d-2}m,\\
 Q&=&\frac{\sqrt{2(d-2)(d-3)}}{8\pi}\omega_{d-2}q,
\end{eqnarray}
with $\omega_{d}=2\pi^{(d+1)/2}/\Gamma((d+1)/2)$ the volume of the unit $d$ sphere.  The black hole temperature, entropy, electric potential, and thermodynamic
volume are given by
\begin{eqnarray}
 T&=&\frac{16\pi r_{h}^2
      \left(P-\frac{2\pi Q^2 r_{h}^{4-2 d}}{\omega_{d-2}^2}\right)
     +(d-5)d+6}{4\pi (d-2) r_{h}},\label{tep}\\
 S&=&\frac{\omega_{d-2}r_{h}^{d-2}}{4},\quad
 \Phi=\frac{4\pi Q r_{h}^{3-d}}{(d-3)\omega},\quad
 V=\frac{\omega_{d-2} r_{h}^{d-1}}{d-1}.
\end{eqnarray}
The Gibbs free energy $G=H-TS$ reads
\begin{eqnarray}
 G=\frac{\omega_{d-2}}{16\pi}\bigg(r_{h}^{d-3}
       -\frac{16\pi Pr_{h}^{d-1}}{(d-1)(d-2)}
       +\frac{32(2d-5)\pi^{2}Q^{2}r_{h}^{3-d}}{(d-2)(d-3)\omega_{d-2}^{2}}\bigg).\label{gbs}
\end{eqnarray}
We plot in Fig. \ref{Pgtq} the characteristic swallow tail behavior of the Gibbs free energy with fixed charge $Q$ and pressure $P$ in the $G-T$ plane. At the point $N$, it can be seen that the Gibbs free energy of the charged black hole vanishes. Note that the Gibbs free energy for AdS space also vanishes. Hence it seems that at that point the black hole phase and the AdS space phase can coexist. However, the concept that ``empty charged space" is unphysical. Thus we could not compare these two phases with fixed charge together, and the AdS phase will not be considered in the following discussion.

After excluding the AdS space phase, the system will first go along the small black hole branch with the increasing of $T$ until the point $M$ is arrived, where the small and large black holes have the same Gibbs free energy, and the small and large black holes can coexist at that point. With the temperature further increasing, the system will prefer the large black hole with lower Gibbs free energy over the small and intermediate black holes. That is the small/large black hole phase transition.

%%%%%%%%%%%%%%%%%%%%%%%%%%%%%%%%%%%%%%%%%%%%%%%%%%%%%%%%%%%%%%%%%%%%%
%%%%%%%%%%%%%%%%%%%%%%%%%%%%%%%%%%%%%%%%%%%%%%%%%%%%%%%%%%%%%%%%%%%%%%%%%

In the extended phase space, the small/large black hole phase transition was well studied in Ref. \cite{Kubiznak}. This phase transition is reminiscent of the liquid/gas phase transition of the vdW fluid. This coexistence curve has positive slope everywhere and terminates at the critical point $(P_{c}, T_{c}, V_{c})$, above which the small and large black hole phases cannot be clearly distinguished. The critical point is determined by
\begin{equation}
 \bigg(\frac{\partial P}{\partial V}\bigg)_{T}=0,\;\;
 \bigg(\frac{\partial^{2} P}{\partial V^{2}}\bigg)_{T}=0.
\end{equation}
By solving the above two equations, we get the critical point:
\begin{eqnarray}
 V_{c}&=&\frac{\omega_{d-2}}{d-1}\bigg(\frac{d-2}{4}\bigg)^{d-1}v_{c}^{d-1},\quad
 T_{c}=\frac{4(d-3)^{2}}{(d-2)(2d-5)\pi v_{c}},\nonumber\\
 P_{c}&=&\frac{(d-3)^{2}}{(d-2)^{2}\pi v_{c}^{2}},\quad
 G_{c}=\sqrt{\frac{2(2d-5)}{d-3}}\frac{Q}{d-1}, \label{critical}
\end{eqnarray}
with the critical specific volume $v_{c}=\frac{d-2}{4}\left(q^{2}(d-2)(2d-5)\right)^{1/(2d-6)}$.

Here we would like to give a brief discussion of the thermodynamic quantities and phase transition in the reduced parameter space. The reduced thermodynamic quantity corresponding to $A$ can be defined as
\begin{equation}
 \tilde{A}=\frac{A}{A_{c}},
\end{equation}
with $A_{c}$ being its critical value. Then the reduced temperature and the Gibbs free energy can be expressed as
\begin{eqnarray}
 \tilde{T}&=&-\frac{\tilde{V}^{\frac{1}{1-d}}
   \left(-2 d^2-(d-3) (2 d-5) \tilde{P} \tilde{V}^{\frac{2}{d-1}}+\tilde{V}^{-\frac{2 (d-3)}{d-1}}+9
   d-10\right)}{4 (d-3) (d-2)},\label{tp}\\
   \tilde{G}&=&\frac{-(d-3)^2 \tilde{V}\tilde{P}+(d-1)
   \tilde{V}^{\frac{3-d}{d-1}}+(d-2)
   (d-1)\tilde{V}^{\frac{d-3}{d-1}}}{4(d-2)},
\end{eqnarray}
from which we can see that $\tilde{T}$ and $\tilde{G}$ are independent of the black hole charge $Q$. Using the fact that the phase transition point is determined by the swallow tail behavior of the Gibbs free energy, we can get the natural result that the phase transition in the reduced parameter space is charge independent. This can also be found from the equal area law. For example, the equal area law on the $P-V$ line is described by Eq. (\ref{vp0}) with fixed $Q$ and $T$. A simple calculation shows that $\int VdP=V_{c}P_{c}\int \tilde{V}d\tilde{P}=0$. Then we have the equal area law in the reduced parameter space,
\begin{eqnarray}
 \int_{\tilde{P}_{A}}^{\tilde{P}_{A}'} \tilde{V}d\tilde{P}=0.
\end{eqnarray}
Since the reduced temperature in (\ref{tp}) is charge independent, the phase transition point determined by constructing the equal area law on the (reduced) isotherm is also charge independent. It is also worthwhile to point out that, in the reduced parameter space, all of the reduced physical quantities have a value of order 1. A word to conclusion is that there is a charge independent coexistence curve for any $d$.

\subsection{($d=4$)-dimensional charged AdS black hole}

\begin{figure}
\center{\subfigure[]{\label{Pqt4}
\includegraphics[width=8cm,height=6cm]{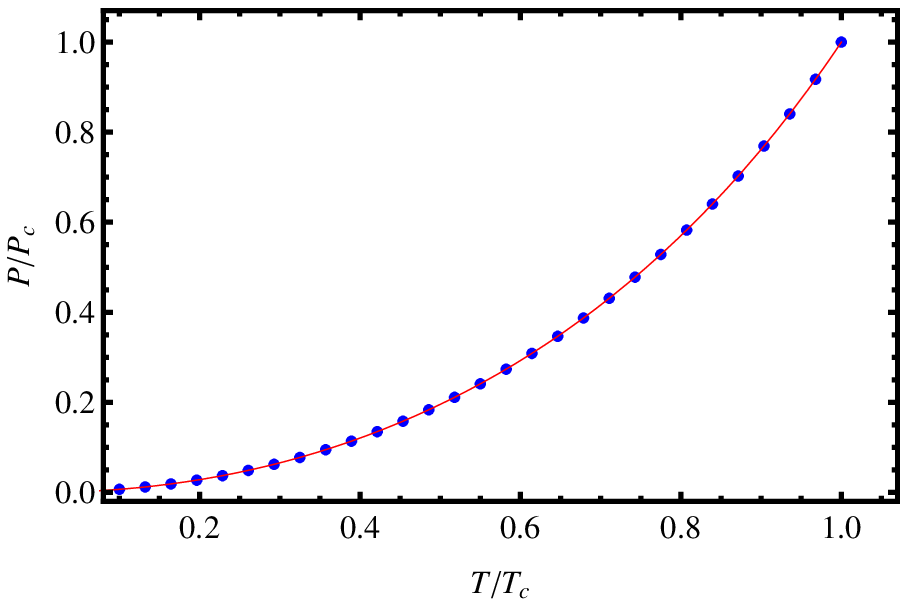}}
\subfigure[]{\label{Pqt4b}
\includegraphics[width=8cm,height=6cm]{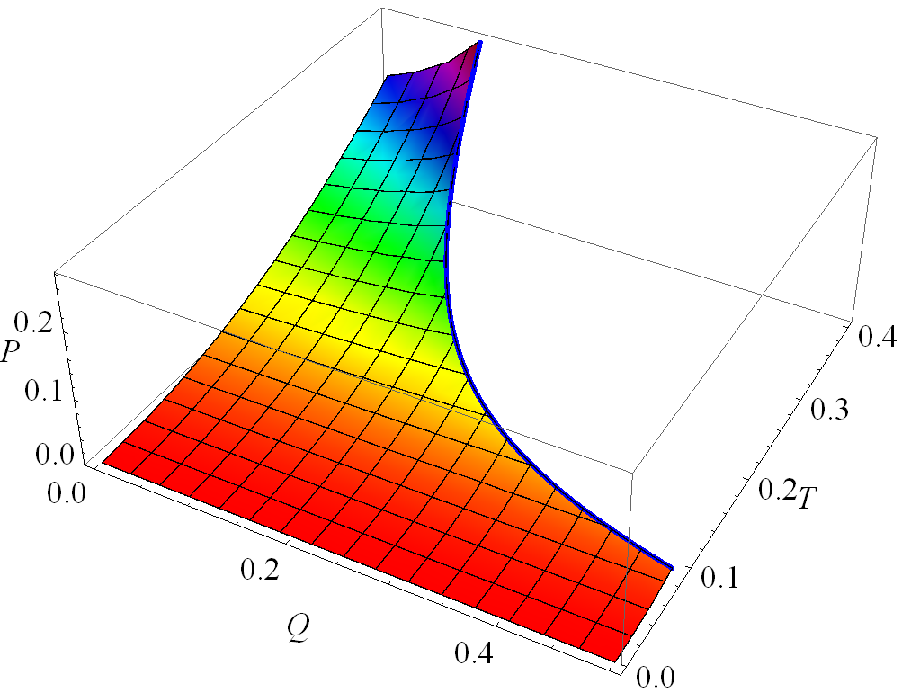}}}
\caption{(a) Coexistence curve of the four-dimensional charged AdS black hole in the reduced parameter space. The discrete points denote the numerical values and the solid line is obtained from the fitting formula Eq. (\ref{coex4}). (b) Coexistence surface of the four-dimensional charged AdS black hole as functions of $Q$ and $T$. The boundary marked with the blue line corresponds to the critical points.}
\end{figure}

For the case of $d=4$, the values of these critical parameters reduce to
\begin{equation}
 V_{c}=8\sqrt{6}\pi Q^{3},\quad
 T_{c}=\frac{\sqrt{6}}{18\pi Q},\quad
 P_{c}=\frac{1}{96\pi Q^{2}},\quad
 G_{c}=\frac{\sqrt{6}}{3}Q.\label{crpo4}
\end{equation}
And the reduced temperature and Gibbs free energy are
\begin{eqnarray}
 \tilde{T}&=&\frac{3 \tilde{P} \tilde{V}^{4/3}+6
   \tilde{V}^{2/3}-1}{8 \tilde{V}},\\
 \tilde{G}&=&\frac{1}{8} \left(-\tilde{P}\tilde{V}+6
   \sqrt[3]{\tilde{V}}+\frac{3}{\sqrt[3]{\tilde{V}}}\right).
\end{eqnarray}
According to the swallow tail behavior of $\tilde{G}$ or the equal area law, the phase diagram in the $\tilde{P}-\tilde{T}$ plane (or the coexistence curve) can be obtained. According to our above analysis, the coexistence curve is charge independent. On the other hand, as shown in many works, this coexistence curve is of the vdW type, and it has positive slope everywhere and terminates at the critical point. Thus, we can fit it using the polynomial of $\tilde{T}$. Combined with the calculated values, the parametrization form of it reads
\begin{eqnarray}
 \tilde{P}&=&0.666902\tilde{T}^2+0.175830 \tilde{T}^3            +0.127273 \tilde{T}^4-0.230638 \tilde{T}^5\nonumber\\
   &+&0.795846 \tilde{T}^6-1.36972 \tilde{T}^7+1.47494 \tilde{T}^8-0.867209
   \tilde{T}^9+0.226773 \tilde{T}^{10},\quad \tilde{T}\in(0,1).  \label{coex4}
\end{eqnarray}
We plot the numerical values (the discrete points) and the fitting formula (the solid line) of the coexistence curve in Fig. \ref{Pqt4}. It is quite clear that the numerical values and the fitting formula match well with each other. Inverting $(\tilde{T}, \tilde{P})$ to $(T, P)$, we will obtain a coexistence surface, which is charge dependent, and we show it in Fig. \ref{Pqt4b}.

%%%%%%%%%%%%%%%%%%%%%%%%%%%%%%%%%%%%%%%%%%%%%%%%%%%%%%%%%%%%%%%%%%%%%
%%%%%%%%%%%%%%%%%%%%%%%%%%%%%%%%%%%%%%%%%%%%%%%%%%%%%%%%%%%%%%%%%%%%%%%%%

%%%%%%%%%%%%%%%%%%%%%%%%%%%%%%%%%%%%%%%%%%%%%%%%%%%%%%%%%%%%%%%%%%%%%
\begin{figure}
\center{\subfigure[]{\label{Ppv}
\includegraphics[width=8cm,height=6cm]{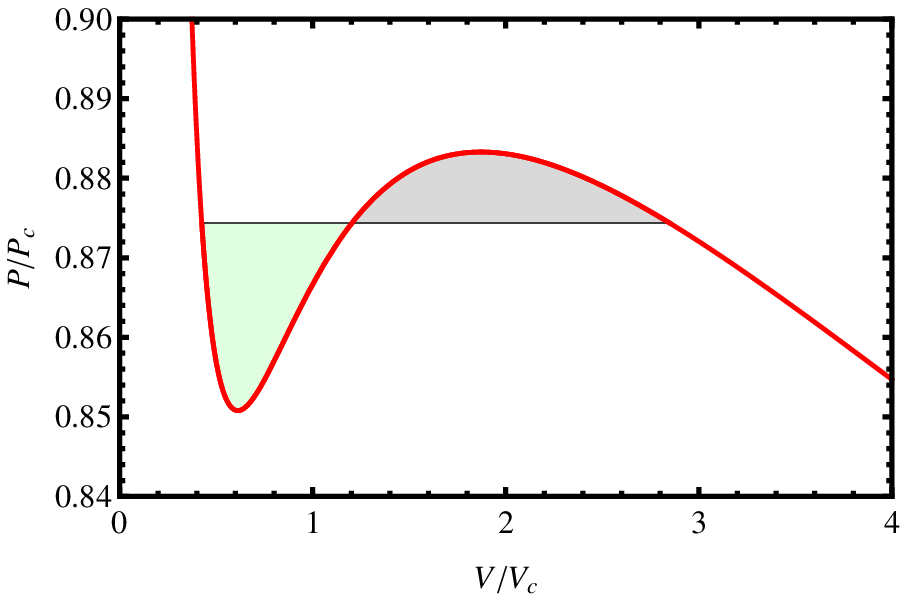}}
\subfigure[]{\label{Ppv2}
\includegraphics[width=8cm,height=6cm]{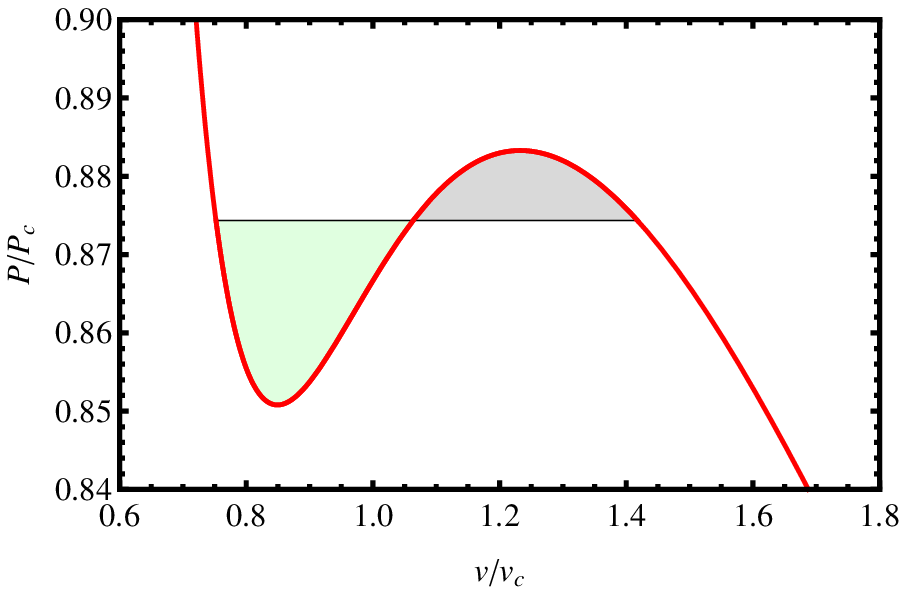}}}\\
\center{\subfigure[]{\label{Pts}
\includegraphics[width=8cm,height=6cm]{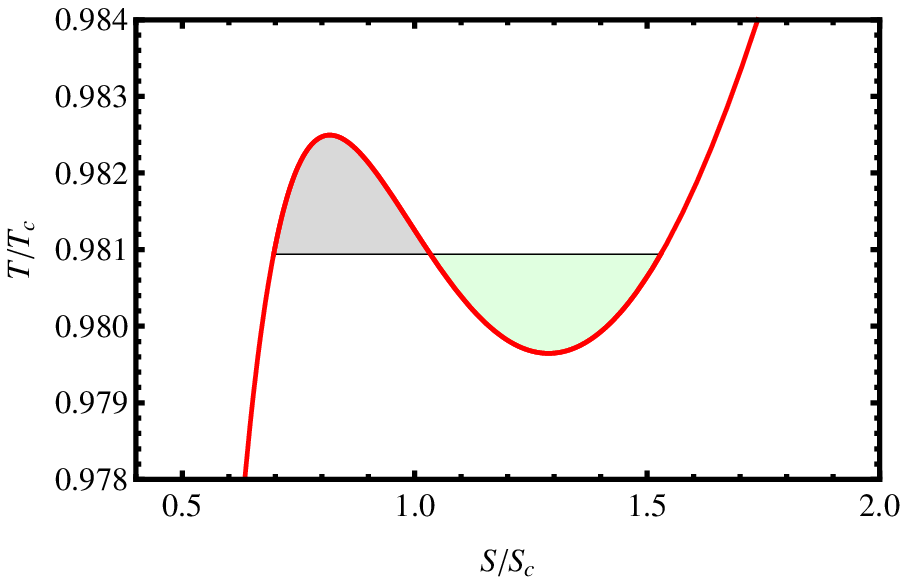}}
\subfigure[]{\label{Pqphi}
\includegraphics[width=8cm,height=6cm]{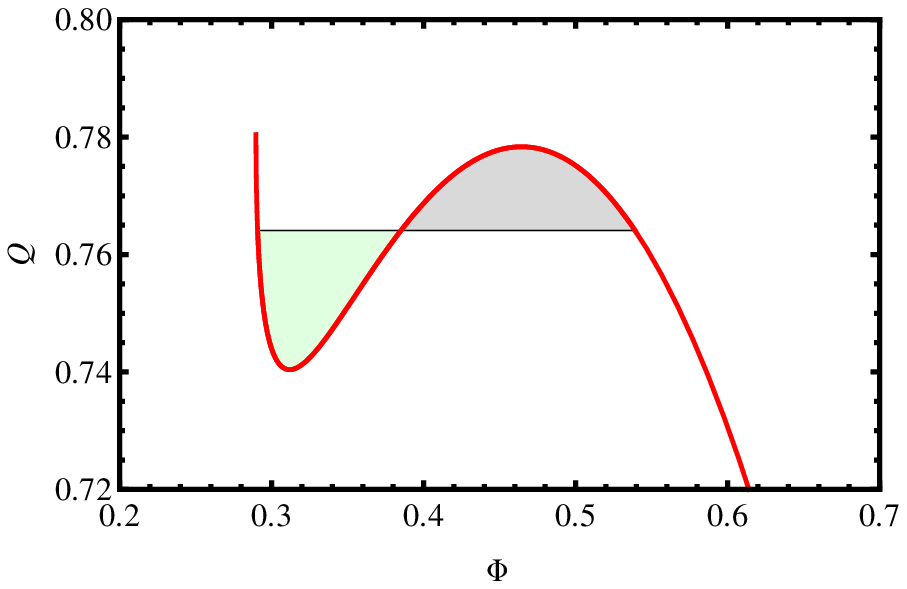}}}
\caption{Area law for the charged AdS black hole. (a) $Q=0.8$, $T/T_{c}=0.95$. Phase transition point at $P/P_{c}=0.8744$. (b) $Q=0.8$, $T/T_{c}=0.95$. Phase transition point at $P/P_{c}=0.8744$. The two areas have a difference of 0.0023. (c) $Q=0.8$, $P/P_{c}=0.95$. Phase transition point at $T/T_{c}=0.9809$. (d) $P=0.005$, $T=0.054$. Phase transition point at $Q=0.7641$.}\label{Parealaw}
\end{figure}
%%%%%%%%%%%%%%%%%%%%%%%%%%%%%%%%%%%%%%%%%%%%%%%%%%%%%%%%%%%%%%%%%%%%%%%%%

With the help of the fitting formula of the coexistence curve (\ref{coex4}), we would like to check the equal area law and the Clapeyron equations obtained in Sec. \ref{dddd}.

The equal area law is shown in Fig. \ref{Parealaw}. These figures show that the three equal area laws discussed above hold. It is also worthwhile to note that the equal area law in the $P-v$ plane fails. From Fig. \ref{Ppv2}, it can be seen that these two areas in the shadow are not equal to each other, and a difference of 0.0023 exists. This supports our result that the equal area law holds only for the isotherm in the $P-V$ plane rather the $P-v$ plane. This result implies a difference between the thermodynamic and specific volumes. On the other hand, the critical point can be equivalently determined by one of the following four conditions: (I) $(\partial_{V} P)_{T,Q}=(\partial^{2}_{V,V} P)_{T,Q}=0$, (II) $(\partial_{v} P)_{T,Q}=(\partial^{2}_{v,v} P)_{T,Q}=0$, (III) $(\partial_{S} T)_{P,Q}=(\partial^{2}_{S,S} T)_{P,Q}=0$, or (IV) $(\partial_{Q} \Phi)_{T,P}=(\partial^{2}_{\Phi,\Phi} Q)_{T,P}=0$.

\begin{figure}
\center{\subfigure[$Q=2$]{\label{Pdeltaa}
\includegraphics[width=8cm,height=6cm]{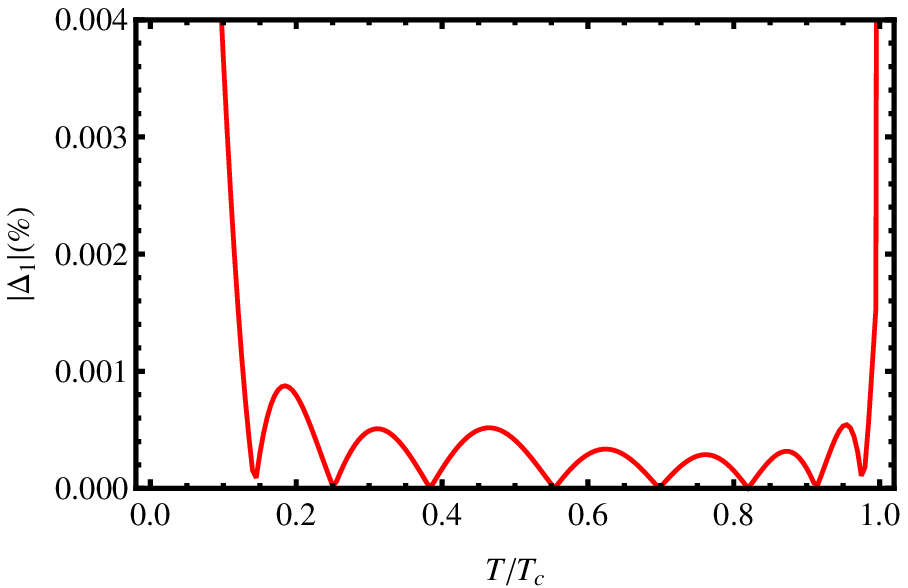}}
\subfigure[$T=0.1$]{\label{Pdeltab}
\includegraphics[width=8cm,height=6cm]{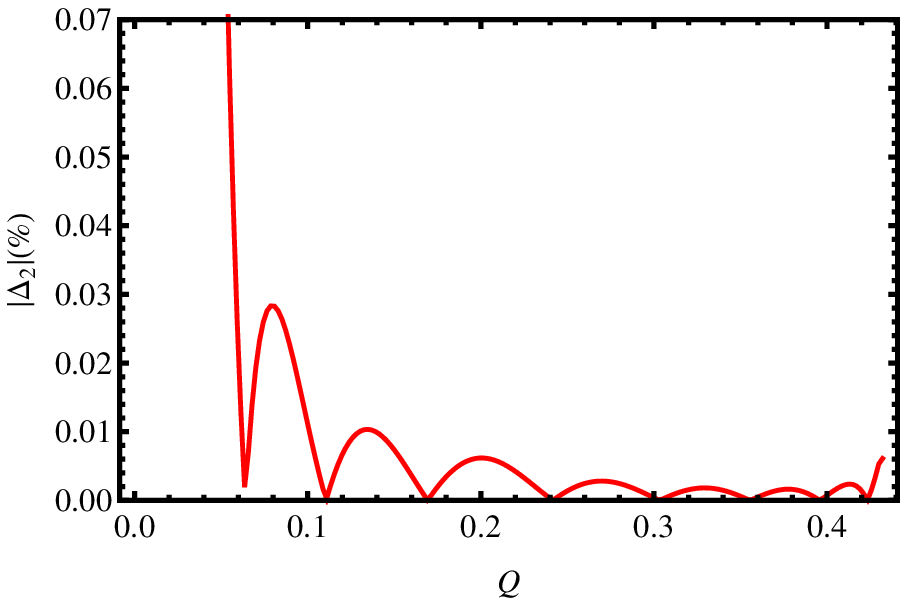}}}
\caption{Behaviors of the deviations $|\Delta_{1}|$ and $|\Delta_{2}|$. }\label{Pdelta}
\end{figure}

In Ref. \cite{Dolan3}, the author pointed out that the Clapeyron equation (\ref{Clapeyron}) holds for the black hole/AdS phase transition in the nonrotating and rotating black hole systems. In the following, we will numerically check Eqs. (\ref{Clapeyron})-(\ref{Clapeyron4}) for the small/large black hole phase transition using the parametrization form of the coexistence curve Eq. (\ref{coex4}). In order to make a clear comparison, we define two relative deviations
\begin{eqnarray}
 \Delta_{1}=\frac{\left(\frac{dP}{dT}\right)_{Q}-\left(\frac{\Delta S}{\Delta V}\right)}{\left(\frac{\Delta S}{\Delta V}\right)}, \quad \Delta_{2}=\frac{\left(\frac{dQ}{dP}\right)_{T}-\left(-\frac{\Delta V}{\Delta \Phi}\right)}{\left(-\frac{\Delta V}{\Delta \Phi}\right)},
\end{eqnarray}
where $\left(\frac{\Delta S}{\Delta V}\right)$ and $\left(-\frac{\Delta V}{\Delta\Phi}\right)$ are numerical values, and $\left(\frac{dP}{dT}\right)_{Q}$ and $\left(\frac{dQ}{dP}\right)_{T}$ are calculated with Eq. (\ref{coex4}). $\Delta_{1}$ measures the deviation of $\left(\frac{dP}{dT}\right)_{Q}$ from $\left(\frac{\Delta S}{\Delta V}\right)$, and $\Delta_{2}$ the deviation of $\left(\frac{dQ}{dP}\right)_{T}$ from $\left(-\frac{\Delta V}{\Delta\Phi}\right)$. The numerical check in high temperature is listed in Table \ref{tab1}, which shows that the relative deviations $|\Delta_{1}|$ and $|\Delta_{2}|$ are $10^{-7}$ and $10^{-6}$, respectively. They are also shown in Fig. \ref{Pdelta} for $Q=2$ and $T=0.1$, respectively. With a fixed $Q$, $|\Delta_{1}|$ shows an oscillatory behavior. At low $T/T_{c}$, $|\Delta_{1}|$ is about $0.02\%$ and quickly decreases below $0.001\%$ with the temperature increasing. Near the critical value of $\tilde{T}=1$, $|\Delta_{1}|$ blows up, which is caused by the fitting method. $|\Delta_{2}|$ also shares the oscillatory behavior. At small charge $Q$, it is about $0.3\%$, while it decreases to $0.01\%$ at large $Q$. Combining with the above results, we can conclude that the Clapeyron equations (\ref{Clapeyron}) and (\ref{Clapeyron2}) hold among the small/large black hole phase transition of the charged AdS black hole. The third Clapeyron equation (\ref{Clapeyron3}) will naturally hold when we consider the Maxwell's relation $\big(\frac{dP}{dT}\big)_{Q}\big(\frac{dT}{dQ}\big)_{P}\big(\frac{dQ}{dP}\big)_{T}=-1$. It is also straightforward to check that the pressure calculated from Eq. (\ref{Clapeyron4}) is exactly consistent with Eq. (\ref{coex4}).

%%%%%%%%%%%%%%%%%%%%%%%%%%%%%%%%%%%%%%%%%%%%%%%%%%%%%%%%%%%%%%%%%%%%%
%%%%%%%%%%%%%%%%%%%%%%%%%%%%%%%%%%%%%%%%%%%%%%%%%%%%%%%%%%%%%%%%%%%%%%%%%

%%%%%%%%%%%%%%%%%%%%%%%%%%%%%%%%%%%%%%%%%%%%%%%%%%%%%%%%%%%%%%%%%%%%%%%%%%%%
\begin{table}[h]
\begin{center}
\begin{tabular}{ccccccccc}
  \hline\hline
  % after \\: \hline or \cline{col1-col2} \cline{col3-col4} ...
      ($Q$, $T$, $P$) & $r_{h}^{L}$ & $r_{h}^{S}$ & $\frac{\Delta S}{\Delta V}$ & -$\frac{\Delta V}{\Delta \Phi}$& $\left(\frac{dP}{dT}\right)_{Q}$ & $\left(\frac{dQ}{dP}\right)_{T}$ &  $|\Delta_{1}|(10^{-7})$ & $|\Delta_{2}|(10^{-6})$ \\
\hline
  (0.3, 0.12, 0.023037) & 1.49733 & 0.456073 & 0.467637 & 29.8718 & 0.467636 & 29.8720 & 9.73743 & 4.88365\\
  (0.6, 0.06, 0.005759) & 2.99465 & 0.912146 & 0.233818 & 238.975 & 0.233818 & 238.976 & 9.73743 & 4.88365\\
  (0.9, 0.04, 0.002560) & 4.49198 & 1.36822 & 0.155879 & 806.540 & 0.155879 & 806.544 & 9.73743 & 4.88365\\
  (1.2, 0.03, 0.001440) & 5.98931 & 1.82429 & 0.116909 & 1911.80 & 0.116909 & 1911.81 & 9.73743 & 4.88365\\
  (1.5, 0.02, 0.000598) & 10.5214 & 2.01368 & 0.069157 & 8042.98 & 0.069157 & 8043.01 & 3.66340 & 3.97774\\
  (1.8, 0.01, 0.000134) & 25.9191 & 2.07298 & 0.028766 & 91254.4 & 0.028766 & 91251.2 & 30.7503 & 34.4390\\\hline\hline
\end{tabular}
\caption{Values of the thermodynamic quantities and deviations $|\Delta_{1}|$ and $|\Delta_{2}|$ at the phase transition points at high temperature. $r_{h}^{L}$ and $r_{h}^{S}$ correspond to the radius of the large and small black holes.}\label{tab1}
\end{center}
\end{table}
%%%%%%%%%%%%%%%%%%%%%%%%%%%%%%%%%%%%%%%%%%%%%%%%%%%%%%%%%%%%%%%%%%%%%%%%%%%%%%

In summary, by adopting the numerical check and the fitting formula of the coexistence curve, we show that the equal area laws and Eqs. (\ref{Clapeyron})-(\ref{Clapeyron4}) derived above hold among the small/large black hole phase transition of the charged AdS black hole.

\subsection{($d>4$)-dimensional charged AdS black hole}

\begin{figure}
\includegraphics[width=8cm,height=6cm]{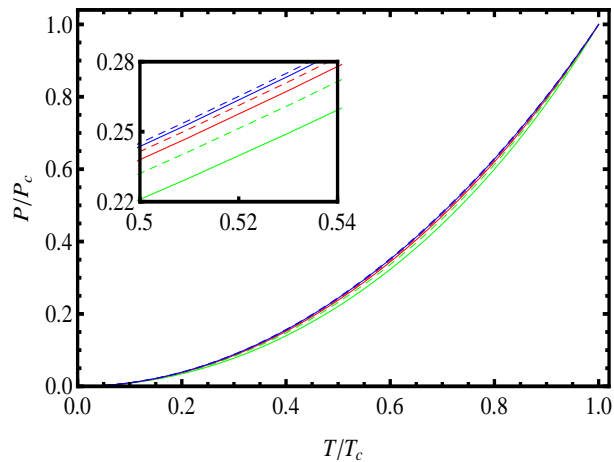}
\caption{The coexistence curves of the charged AdS black holes in the reduced parameter space with $d=5,6,7,8,9,10$ from bottom to top.}\label{Ppptt}
\end{figure}

Similar to the four-dimensional charged AdS black hole, the small/large black hole phase transition also exists in higher-dimensional spacetime. The critical point is given in Eq. (\ref{critical}). After the numerical checking, the Clapeyron equations also hold. The parametrization form of the coexistence curve can also be obtained with fitting the numerical data, which are also charge independent from our above analysis in the reduced parameter space:
\begin{eqnarray}
 \tilde{P}=\sum^{10}_{i=0} a_{i}\tilde{T}^{i},\quad\tilde{T}\in(0,1), \label{tildePTd}
\end{eqnarray}
where $a_{i}$ are the fitting coefficients, and the result shows $a_{0}=a_{1}=0$. For $d=5\sim 10$, we list $a_{i}$ ($i=2\sim10$) in Table \ref{tab2}. The coexistence curves are shown in the reduced parameter space in Fig. \ref{Ppptt}. For a different $d$, we find that the curves are almost the same.

%%%%%%%%%%%%%%%%%%%%%%%%%%%%%%%%%%%%%%%%%%%%%%%%%%%%%%%%%%%%%%%%%%%%%%%%%%%%
\begin{table}[h]
\begin{center}
\begin{tabular}{cccccccccccc}
  \hline\hline
  % after \\: \hline or \cline{col1-col2} \cline{col3-col4} ...
  $d$& $a_{2}$& -$a_{3}$& $a_{4}$& -$a_{5}$& $a_{6}$& -$a_{7}$& $a_{8}$& -$a_{9}$& $a_{10}$\\\hline
  5&0.853443&0.002568&0.151056&0.168247&0.470950&0.789347&0.866793& 0.523890&0.141821\\
  6&0.918039&0.005929&0.041750&0.242304&0.331168&0.310086&0.088335& 0.061263&0.046188\\
  7&0.947497&0.012432&0.093990&0.371470&0.770814&1.089190&0.845138& 0.323729&0.034370\\
  8&0.963482&0.013828&0.107795&0.441760&1.049480&1.544330&1.247598& 0.512227&0.070762\\
  9&0.974263&0.010166&0.087897&0.404075&1.096790&1.825630&1.898470& 1.090380&0.272828\\
 10&0.980271&0.007229&0.072804&0.378345&1.137690&2.064940&2.248680&
  1.32137&0.332741\\\hline\hline
\end{tabular}
\caption{Values of the coefficients $a_{i}$ ($i=2\sim10$) in the fitting formula of the coexistence curves.}\label{tab2}
\end{center}
\end{table}
%%%%%%%%%%%%%%%%%%%%%%%%%%%%%%%%%%%%%%%%%%%%%%%%%%%%%%%%%%%%%%%%%%%%%%%%%%%%%%

%%%%%%%%%%%%%%%%%%%%%%%%%%%%%%%%%%%%%%%%%%%%%%%%%%%%%%%%%%%%%%%%%%%%%
%%%%%%%%%%%%%%%%%%%%%%%%%%%%%%%%%%%%%%%%%%%%%%%%%%%%%%%%%%%%%%%%%%%%%%%%%

\section{Discussions and conclusions}
\label{Conclusion}

In this paper, we studied the thermodynamic properties of the charged AdS black hole along the coexistence curve of the small and large black hole phases. We first reviewed the classical $P-V$ oscillatory behavior, which implies the phase transition taking place. By means of the Maxwell's equal area law, the unphysical branch of the system can be excluded, and the phase transition point can also be determined. For the charged AdS black hole, we showed that such $P-V$ oscillatory behavior is also obvious in the charge-electric potential ($Q-\Phi$) and $T-S$ planes. At each oscillatory line, the equal area law is useful for finding the phase transition point, and which is exactly consistent with the analysis of the Gibbs free energy. It is worthwhile to point out that the oscillatory behavior also appears in the $P-v$ plane. However, its equal area law fails to produce the correct phase transition point, which reminds us of the difference between the thermodynamic and specific volumes.

Employing the first law, we get three Clapeyron equations at the coexistence curve. The first one is the conventional Clapeyron equation with fixed charge, and the other two are its generalizations with, respectively, fixed pressure and temperature.

Before carrying out the numerical check of the Clapeyron equations, we examined the thermodynamic quantities in the reduced parameter space. The reduced temperature $\tilde{T}$ and the Gibbs free energy $\tilde{G}$ are found to be charge independent, which will produce a charge independent phase transition, as well as the coexistence curve. The charge independent property of the coexistence curve can also be found from the state equation expressed with the reduced parameters; see (\ref{tp}). As a consequence, it is a universal property for the charged AdS black hole of any dimension $d$. The numerical calculation also confirms this result. However, after converting to the parameters $P$ and $T$, the coexistence surface is obviously charge dependent. Making use of the fitting formula, we checked that these Clapeyron equations hold exactly when the black hole system varies along the coexistence curve. For example, at low temperature, the deviation is of $0.02\%$, and it quickly decreases to $0.001\%$ for high temperature. A simple check shows that our fitting formulas of the coexistence curve agree with the calculated values to $10^{-7}$.

At last, we would like to make a few comments. Recent studies mainly focus on finding the vdW-type phase transition, reentrant phase transitions or thermodynamic properties near the critical point, while only a few papers address the properties of the black hole system going along or crossing the coexistence curve. We argue that studying the physics near the coexistence curve may help us to reveal the information of the microscopic structure of a thermodynamic system, which is still unclear from the gravity side. So we hope that this study of the phase transition can provide us with some information about the microscopic structure of a gravitational system from the thermodynamic viewpoint. On the other hand, we presented in this paper the highly accurate fitting formulas of the coexistence curves [see Eqs. (\ref{coex4}) and (\ref{tildePTd}) as well as Table \ref{tab2}]. Further research, such as the exploration of the thermodynamic geometry and examination of the quasinormal modes  along the coexistence curves can be easily carried out based on them. On the other hand, seeking the fitting formula of the coexistence curve for the AdS black hole systems with the rich phase structure, such as the higher-dimensional Kerr-Ads black hole system or those in modified gravity, is also an interesting issue.

\section*{Acknowledgements}
We thank R. B. Mann for providing the Maple file and for the valuable discussions on the equal area law. The authors would especially like to thank the anonymous referees whose comments helped us greatly in improving the original manuscript. This work was supported by the National Natural Science Foundation of China (Grants No. 11205074 and No. 11375075).

\end{document}